\documentclass{article}
\def\Ref#1{(\ref{#1})}
\usepackage{amsmath}
\usepackage{amssymb}
\usepackage{cite}
\def\d{{\rm d}}
\begin{document}
\begin{titlepage}
\noindent{\large\textbf{Static- and dynamical-phase transition in
one-dimensional reaction-diffusion systems with boundaries}}

\vskip 2 cm

\begin{center}{Mohammad~Khorrami$^a${\footnote
{mamwad@iasbs.ac.ir}} \& Amir~Aghamohammadi$^b${\footnote
{mohamadi@azzahra.ac.ir}}} \vskip 5 mm

\textit{ $^a$ Institute for Advanced Studies in Basic Sciences,
             P.~O.~Box 159,\\ Zanjan 45195, Iran. }

\textit{ $^b$ Department of Physics, Alzahra University,
             Tehran 19834, Iran. }
\end{center}

\begin{abstract}
\noindent A general system of particles (of one or several
species) on a one dimensional lattice with boundaries is
considered. Two general behaviors of such systems are
investigated. The stationary behavior of the system, and the
dominant way of the relaxation of the system toward its stationary
system. Bases on the first behavior, static phase transitions
(discontinuous changes in the stationary profiles of the system)
are studied. Based on the second behavior, dynamical phase
transitions (discontinuous changes in the relaxation-times of the
system) are studied. The investigation is specialized on systems
in which the evolution equation of one-point functions are closed
(the \textit{autonomous} systems).
\end{abstract}
%
\end{titlepage}
\section{Introduction}
The study of the reaction-diffusion systems, has an attractive
area. A reaction-diffusion system consists of a collection of
particles (of one or several species) moving and interacting with
each other with specific probabilities (or rates in the case of
continuous time variable). In the so called exclusion processes,
any site of the lattice the particles move on, is either vacant or
occupied by one particle. The aim of studying such systems, is of
course to calculate the time evolution of such systems. But to
find the complete time evolution of a reaction-diffusion system,
is generally a very difficult (if not impossible) task.

Reaction-diffusion system have been studied using various methods:
analytical techniques, approximation methods, and simulation. The
success of the approximation methods, may be different in
different dimensions, as for example the mean field techniques,
working good for high dimensions, generally do not give correct
results for low dimensional systems. A large fraction of
analytical studies, belong to low-dimensional (specially
one-dimensional) systems, as solving low-dimensional systems
should in principle be easier
\cite{ScR,ADHR,KPWH,HS1,PCG,HOS1,HOS2,AL,AKK,RK,RK2,AKK2,AM1}.

Various classes of reaction-diffusion systems are called
exactly-solvable, in different senses. In \cite{AA} and
\cite{RK3}, integrability means that the $N$-particle conditional
probabilities' S-matrix is factorized into a product of 2-particle
S-matrices. This is related to the fact that for systems solvable
in this sense, there are a large number of conserved quantities.
In \cite{BDb,BDb1,BDb2,BDb3,Mb,HH,AKA,KAA,MB,AAK}, solvability
means closedness of the evolution equation of the empty intervals
(or their generalization).

Consider a reaction-diffusion system (on a lattice) with open
boundaries. By open boundaries, it is meant that in addition to
the reactions in the bulk of the lattice, particles at the
boundaries do react with some external source. A question is to
find the possible phase transitions of the system. By phase
transition, it is meant a discontinuity in some behavior of the
system with respect to its parameters. Such discontinuities, may
arise in two general categories: in the stationary (large time)
profiles of the system, and in the time constants determining the
evolution os the system. In the first case, static phase
transitions are dealt with; in the second case, dynamical phase
transitions. As mentioned before, the task of finding the complete
evolution of a general reaction-diffusion system is very difficult
(if not impossible). So our studies are limited to a certain class
of reaction-diffusion systems, and in them to certain properties.
To be specific, we deal with one-dimensional systems for which the
evolution equation of one-point functions (probabilities of
finding a particle of a certain kind on a certain point) is closed
(\textit{autonomous} systems). For these systems, we only consider
the final (stationary) profile of the one-point function, and the
relaxation-time (the most significant one) towards the stationary
profile.

In \cite{GS}, a ten-parameter family of one-species
reaction-diffusion processes with nearest-neighbor interaction was
introduced, for which the evolution equation of $n$-point
functions contains only $n$- or less- point functions. The average
particle-number in each site has been obtained exactly for these
models. In \cite{AAMS,SAK}, this has been generalized to
multi-species systems and more-than-two-site interactions. In
\cite{MA1,AM2,MAM,MA2}, the phase structure of some classes of
single- or multiple-species reaction-diffusion systems have been
investigated. These investigations were bases on the one-point
functions of the systems.

The scheme of the present article is as follows. In section 2, the
conditions for the most general multi-species reaction-diffusion
models with nearest-neighbor interactions, to be solvable, or
autonomous, are obtained. In section 3, autonomous single-species
reaction-diffusion models with boundaries have been investigated.
It is shown that changing the bulk rates may lead to a static
phase transition, while changing the bulk or boundary rates may
lead to a dynamical phase transition. In section 4, as an example,
an asymmetric generalization of the Glauber model at zero
temperature is introduced. This system exhibits both static and
dynamical phase transitions. In section 5, as an example of
systems with more than two-site interaction, Glauber model at
finite temperature has been considered. With specific boundary
interactions, the Glauber model with open boundaries is also
autonomous. It is shown that although this system does not show a
static phase transition, it does exhibit a dynamical phase
transition. In the last section, as an example of multi-species
reaction-diffusion systems, an asymmetric generalization of the
zero temperature Potts model is introduced. This system exhibits
both a static and a dynamical phase transition.
\section{Autonomous reaction-diffusion systems,
with nearest-neighbor interactions} In \cite{AAMS}, multi-species
reaction diffusion with nearest-neighbor interactions on a
one-dimensional lattice are studied, and criteria are obtained
that such systems be autonomous, meaning that the evolution
equation for the one-point functions be closed. The summary of the
work follows.

Let the Hamiltonian of the system be
\begin{equation}\label{1}
{\cal H}=\sum_{i=1}^{L-1}H_{i,i+1},
\end{equation}
where
\begin{equation}\label{2}
H_{i,i+1}:=\underbrace{1\otimes\cdots\otimes 1}_{i-1}\otimes
H\otimes\underbrace{1\otimes\cdots\otimes 1}_{L-i-1}.
\end{equation}
$1$ is the $(p+1)\times(p+1)$ unit matrix, and $H$ is a
$(p+1)\times(p+1)$ matrix (for a $p$ species system), the
nondiagonal elements of which are nonnegative. $H$ also satisfies
\begin{equation}\label{3}
\mathbf{s}\, H=0,
\end{equation}
where $\mathbf{s}$ is a covector with all of its components equal
to 1:
\begin{equation}\label{4}
s_\alpha=1.
\end{equation}
The number operator of a particle of type $A_\alpha$ in the site
$i$ is denoted by $n^\alpha_i$, where $A_{p+1}$ can be the
vacancy. These number operators are of the form
\begin{equation}\label{5}
n^\alpha_i:=\underbrace{1\otimes\cdots\otimes 1}_{i-1}\otimes
n^\alpha\otimes\underbrace{1\otimes\cdots\otimes 1}_{L-i},
\end{equation}
where $n^\alpha$ is a diagonal $(p+1)\times(p+1)$ matrix, the only
nonzero element of which is the element $\alpha$ of the diagonal.
It is clear that the vector $\mathbf{n}$ whose components are
$n^\alpha$, satisfies
\begin{equation}\label{6}
\mathbf{s}\,\mathbf{n}=1,\qquad\hbox{ or }\qquad
s_\alpha\,n^\alpha=1.
\end{equation}

The Hamiltonian $\cal H$ is the generator of time translation, by
which it is meant that
\begin{equation}\label{7}
\frac{\d}{\d t}\mathbf{P}(t)={\cal H}\,\mathbf{P}(t),
\end{equation}
where $\mathbf{P}$ is a the vector in the $(p+1)^L$ dimensional
space, the component $P^{\alpha_1\cdots \alpha_L}$ of which is the
probability of finding a particle of type $A_{\alpha_1}$ in site
$1$, etc. Clearly, the components of this vector are nonnegative
and satisfy
\begin{equation}\label{8}
\mathbf{S}\,\mathbf{P}=1,
\end{equation}
where
\begin{equation}\label{9}
\mathbf{S}:=\underbrace{\mathbf{s}\otimes\cdots\otimes\mathbf{s}}_{L}.
\end{equation}
The fact that $\mathbf{P}$ should have these properties for all
times, means that the nondiagonal elements of $\cal{H}$ should be
nonnegative and that the action of $\cal H$ (from right) on
$\mathbf{S}$ should be zero. But the nonnegativity of the
nondiagonal elements of $H$, and the condition \Ref{3}, are
sufficient for these two conditions on $\cal H$ to be satisfied.

The expectation value of any operator $\cal O$ is
\begin{equation}\label{10}
\langle{\cal O}\rangle=\mathbf{S}\,{\cal O}\,\mathbf{P}.
\end{equation}
This is also true for $n^\alpha_i$, from which it is seen that
\begin{equation}\label{11}
\frac{\d}{\d t}\langle n^\alpha_i\rangle=\mathbf{S}\,n^\alpha_i
\,{\cal H}\,\mathbf{P}.
\end{equation}
Now the question is, under what conditions the right-hand side of
\Ref{11} can be written in terms of the one-point functions
$n^\beta_j$'s?

To answer this, one notes that $n_i^\alpha$ commutes with all of
terms in $\cal H$, except possibly with $H_{i-1,i}$ and
$H_{i,i+1}$. Using
\begin{equation}\label{12}
\mathbf{s}\otimes\mathbf{s}\,(\mathbf{a}\,\mathbf{n})\otimes
(\mathbf{b}\,\mathbf{n})\, H=(\mathbf{a}\otimes\mathbf{b}\,
H)_{\alpha\beta}\,\mathbf{s}\otimes\mathbf{s}\,
\mathbf{n}^\alpha\otimes\mathbf{n}^\beta,
\end{equation}
which is true for any two covectors $\mathbf{a}$ and $\mathbf{b}$,
and using \Ref{6}, it is seen that
\begin{equation}\label{13}
\frac{\d}{\d t}\langle n^\alpha_i\rangle=s_\gamma\,
H^{\gamma\alpha}{}_{\mu\nu}\,\langle n^\mu_{i-1}\, n^\nu_i\rangle+
s_\gamma\, H^{\alpha\gamma}{}_{\mu\nu}\,\langle n^\mu_i\,
n^\nu_{i+1}\rangle.
\end{equation}
The right hand side is expressed in terms of two-point functions
not one-point functions. However, from \Ref{6} it is seen that the
two-point functions are not independent. In order that an
expression $B_{\mu\nu}\,n^\mu_i\,n^\nu_j$ be expressible in terms
of the first power of number operators, it is necessary and
sufficient that
\begin{equation}\label{14}
B_{\mu\nu}={}_1 B_\mu\, s_\nu+s_\mu\, {}_2 B_\nu.
\end{equation}
So the necessary and sufficient condition for the system to be
autonomous, is that
\begin{equation}\label{15}
{}^i H^\alpha{}_{\mu\nu}={}^i_1 H^\alpha{}_\mu\, s_\nu+s_\mu\,
{}^i_2 H^\alpha{}_\nu,
\end{equation}
where
\begin{align}\label{16}
{}^1 H^\alpha{}_{\mu\nu}&:=s_\gamma\,
H^{\alpha\gamma}{}_{\mu\nu},\nonumber\\
{}^2 H^\alpha{}_{\mu\nu}&:=s_\gamma\, H^{\gamma\alpha}{}_{\mu\nu}.
\end{align}
Note that even if \Ref{15} is satisfied, ${}^i_j H$'s are not
uniquely determined through which. One can change them like
\begin{equation}\label{17}
{}^i_j H^\alpha{}_\mu\to {}^i_j H^\alpha{}_\mu+ {}^i_j w^\alpha\,
s_\mu,
\end{equation}
with
\begin{equation}\label{18}
\sum_j {}^i_j\mathbf{w}=0.
\end{equation}
For the simplest case $p=1$ (single-species systems) \Ref{15}
consists of two independent constraints, leaving a ten-parameter
family of autonomous systems out of the twelve-parameter family of
the general systems.

The condition \Ref{15} for a system to be autonomous, can be
easily extended to systems with interaction-ranges longer than the
nearest neighbor. This has been done in \cite{SAK}. The result is
that if the interaction is in a block of $k$ neighboring sites,
then one should have
\begin{equation}\label{19}
{}^i H^\alpha{}_{\beta_1\cdots\beta_k}= {}^i_1
H^\alpha{}_{\beta_1}\,s_{\beta_2}\cdots s_{\beta_k}+\cdots+
+s_{\beta_1}\cdots s_{\beta_{k-1}}\, {}^i_k H^\alpha{}_{\beta_k},
\end{equation}
where
\begin{equation}\label{20}
{}^i H^{\alpha_i}{}_{\beta_1\cdots\beta_k}:=s_{\gamma_1}\cdots
s_{\gamma_{i-1}}\, s_{\gamma_{i+1}}\cdots s_{\gamma_k}\,
H^{\alpha_1\cdots\alpha_k}{}_{\beta_1\cdots\beta_k}.
\end{equation}

Assuming that \Ref{15} holds, one can write the evolution equation
for the one-point functions as
\begin{equation}\label{21}
\frac{\d}{\d t}\langle\mathbf{n}_i\rangle={}^2_1
H\,\langle\mathbf{n}_{i-1}\rangle+ ({}^2_2 H+{}^1_1
H)\,\langle\mathbf{n}_i\rangle +{}^1_2
H\,\langle\mathbf{n}_{i+1}\rangle,\qquad 1<i<L.
\end{equation}
For $i=1$, at the right-hand only the last two terms remain; for
$i=L$, only the first two terms. This is a linear
differential-difference equation for the vectors
$\langle\mathbf{n}\rangle$.
\section{Autonomous single-species reaction-diffusion systems with boundaries}
For a single-species system, the condition \Ref{15} can be
rewritten in the more explicit form
\begin{align}\label{22}
-H^{01}{}_{11}-H^{00}{}_{11}+H^{01}{}_{10}+H^{00}{}_{10}-H^{11}{}_{01}
-H^{10}{}_{01}+H^{11}{}_{00}+H^{10}{}_{00} &=:0,\nonumber\\
-H^{10}{}_{11}-H^{00}{}_{11}-H^{11}{}_{10}-H^{01}{}_{10}+H^{10}{}_{01}
+H^{00}{}_{01}+H^{11}{}_{00}+H^{01}{}_{00} &=:0.
\end{align}
Here the state $A_1$ is an occupied site, while $A_0$ is a
vacancy. Defining
\begin{align}\label{23}
u&:=H^{10}{}_{01}+H^{00}{}_{01}\nonumber\\
v&:=H^{01}{}_{10}+H^{00}{}_{10}\nonumber\\
\bar u&:=H^{10}{}_{11}+H^{00}{}_{11}\nonumber\\
\bar v&:=H^{01}{}_{11}+H^{00}{}_{11}\nonumber\\
w&:=H^{11}{}_{00}+H^{10}{}_{00}\nonumber\\
s&:=H^{11}{}_{00}+H^{01}{}_{00}\nonumber\\
\bar w&:=H^{11}{}_{01}+H^{10}{}_{01}\nonumber\\
\bar s&:=H^{11}{}_{10}+H^{01}{}_{10},
\end{align}
one can write \Ref{22} as
\begin{align}\label{24}
u+s&=\bar u+\bar s\nonumber\\
v+w&=\bar v+\bar w.
\end{align}
The evolution equation of $\langle\mathbf{n}_i\rangle$, for
$1<i<L$ is of the form \Ref{21}. For $i=1$ and $i=L$, only half of
the terms at the right-hand side of \Ref{21} are there. Moreover,
if there is injection and extraction of particles at the end
sites, terms corresponding to these rates should also be added.
Finally, for the single species case, denoting $n^1_i$ by $n_i$,
one can write instead of the vector equation \Ref{21}, a scalar
equation for $n_i$, as $n^1_i+n^0_i=1$. denoting the injection and
extraction rates at the first site by $a$ and $a'$ respectively,
and those at the last site by $b$ and $b'$, the evolution equation
of the one-point functions can be seen to be
\begin{align}\label{25}
\langle\dot n_i\rangle =& -(v+w+u+s)\langle n_i\rangle + (v-\bar
v) \langle n_{i+1}\rangle +(u-\bar u)\langle
n_{i-1}\rangle\nonumber\\
& +w+s,\qquad 1<i<L \nonumber\\
\langle\dot n_1\rangle =& -(v+w)\langle n_1\rangle + (v-\bar v)
\langle n_2\rangle +w+a(1-\langle n_1\rangle )-a' \langle
n_1\rangle\nonumber\\
\langle\dot n_L\rangle =& -(u+s)\langle n_L\rangle + (u-\bar u)
\langle n_{L-1}\rangle +s+b(1-\langle n_L\rangle )-b'\langle
n_L\rangle ,
\end{align}
\subsection{The static phase transition}
The stationary-state to \ref{25} is
\begin{equation}\label{26}
\langle n_j\rangle=C+D_1\, z_1^j+ D_2\, z_2^j,
\end{equation}
where $z_i$'s satisfy
\begin{equation}\label{27}
-r+p\,z_i+q\,z_i^{-1}=0,
\end{equation}
and we take $z_2$ to the root of the above equation with larger
absolute value, and the new parameters $p$, $q$, and $r$ are
defined through
\begin{align}\label{28}
p:=&v-\bar v\nonumber\\
q:=&u-\bar u\nonumber\\
r:=&u+s+v+w=u+s+\bar v+\bar w\nonumber\\
=&\bar u+\bar s+v+w=\bar u+\bar s+\bar v+\bar w,
\end{align}
Noting that the rates (nondiagonal elements of $H$) are
nonnegative, it is seen that
\begin{equation}\label{29}
r\geq |p+q|,|p-q|.
\end{equation}
More over, it is seen that if $r=0$, then all the parameters
introduced in \Ref{23} vanish, and $\langle n_i\rangle$ would be
constant (for $1<i<L$). So apart from this trivial case, $r$ is
positive and in fact can be rescaled to one (by a proper
redefinition of time). Hence, as long as only the stationary
profile of the one-point function is considered, there are only
two-parameters in the system:
\begin{equation}\label{30}
x:=\frac{p}{r},\qquad\hbox{ and }\qquad y:=\frac{q}{r}.
\end{equation}
The physical region corresponding to these parameters is a square
the boundaries of which are $|x\pm y|=1$.

It can be seen that inside the physical square, both of the roots
of \Ref{27} are real, one ($z_1$) between $-1$ and $1$, the other
($z_2$) outside that region. So in the thermodynamic limit
$L\to\infty$,
\begin{align}\label{31}
\langle n_j\rangle &\approx C+D_1\,z_1^j,  &j\ll L\nonumber\\
\langle n_j\rangle &\approx C+D_2\,z_2^{j-L-1},  &L-j\ll L.
\end{align}
$C$, $D_i$'s, and $z_i$'s are smooth functions of rates, and there
is no phase transition.

However, if $|x+y|=1$, then one of the roots of \Ref{27} would be
$x+y$; that is, the absolute value of one of the roots will be 1.
The other roots would be $y/x$. If $|y/x|<1$, then $z_1=1$ and
from \Ref{31} it is seen that near the end of the lattice,
$|\langle n_j\rangle-C|$ is essentially constant. If $|y/x|>1$,
then $z_2=1$ and from \Ref{31} it is seen that near the beginning
of the lattice, $|\langle n_j\rangle-C|$ is essentially constant.
So there is a phase transition at $x=y=\pm 1/2$. This static phase
transition manifests itself as a change in the slope of the
profile of the one-point function, near the end or the beginning
of the lattice.
\subsection{The dynamical phase transition}
The dynamical phase transition is related to the relaxation time
of the system evolving towards its stationary configuration. To
find this relaxation time, one writes the homogeneous part of
\Ref{25} as
\begin{equation}\label{32}
\dot N_i=h_i{}^j\, N_j,
\end{equation}
where $N_i$ is the $\langle n_i\rangle$ minus its stationary
value. Then, one seeks the eigenvalues of $h$. The eigenvalue with
the largest real part, determines the relaxation time:
\begin{equation}\label{33}
\tau=-\frac{1}{\textrm{Re}(E_{\textrm{max}})}.
\end{equation}
The eigenvector equations read
\begin{align}\label{34} E\, X_j&=
-(v+w+u+s)X_j+(v-\bar v)X_{j+1}+(u-\bar u)X_{j-1},\qquad
j\ne 1,L\nonumber\\
E\, X_1&= -(v+w+a+a')X_1+(v-\bar v)X_2,\nonumber\\
E\, X_L&= -(u+s+b+b')X_L+(u-\bar u)X_{L-1}.
\end{align}
A solution for this is
\begin{equation}\label{35}
X_j=\alpha\, z_1^j+\beta z_2^j,
\end{equation}
with
\begin{equation}\label{36}
E=-(v+w+u+s)+(v-\bar v)z_i+(u-\bar u)z_i^{-1},
\end{equation}
and
\begin{align}\label{37}
(v-\bar v)(\alpha z_1^2+\beta z_2^2)-(E+a+a'+v+w)(\alpha z_1+
\beta z_2) &= 0\nonumber\\
(u-\bar u)(\alpha z_1^{L-1}+\beta z_2^{L-1})-(E+b+b'+u+s) (\alpha
z_1^L+\beta z_2^L)&= 0.
\end{align}
One puts $E$ from \Ref{36} in \Ref{36}, and demands \Ref{36} to
have nonzero solutions for $\alpha$ and $\beta$. The result is
\begin{align}\label{38}
[(u-\bar u)+z_1\delta a][(v-\bar v)z_2^{L+1}+z_2^L\delta b]&-
[(u-\bar u)+z_2\delta a]\nonumber\\
&\times [(v-\bar v)z_1^{L+1}+z_1^L\delta b]=0,
\end{align}
where
\begin{align}\label{39}
\delta a&:=a+a'-(u+s),\nonumber\\
\delta b&:=b+b'-(v+w).
\end{align}
Defining the new variables $Z_i$ as
\begin{equation}\label{40}
Z_i:=z_i\sqrt{\left|\frac{v-\bar v}{u-\bar u}\right|},
\end{equation}
it is seen that
\begin{align}\label{41}
E&=-(v+w+u+s)+\sqrt{|(u-\bar u)(v-\bar v)|}[Z_i\,{\rm sgn}(v-\bar
v)+
Z_i^{-1}\,{\rm sgn}(u-\bar u)]\nonumber\\
&=-(v+w+u+s)+{\rm sgn}(v-\bar v)\sqrt{|(u-\bar u)(v-\bar
v)|}(Z_1+Z_2)\nonumber\\
&=-(v+w+u+s)+{\rm sgn}(u-\bar u)\sqrt{|(u-\bar u)(v-\bar
v)|}(Z_1^{-1}+ Z_2^{-1}).
\end{align}
In the thermodynamic limit $L\to\infty$, all unimodular complex
numbers are solutions for $Z_i$. The maximum of the real part of
$E$ depends on whether all of the values of $Z_i$ are (unimodular)
or not. In the first case, the relaxation time is independent of
the injection and extraction rates at the boundaries. This is
called the \textit{fast} phase. In the second case, the relaxation
time does depend on the injection and extraction rates at the
boundaries. This is called the \textit{slow} phase. The terms fast
and slow come from the fact that for fixed bulk reaction rates,
the relaxation time in the fast phase is smaller than that of the
slow phase. Now the physical parameter space can be divided to
three regions, with different behaviors for the relaxation-time:
\begin{equation}\label{42}
\begin{cases}
a+a'<{\cal A},\quad a+a'-(b+b')<{\cal A}-{\cal B},& \hbox{region
I}\\
b+b'<{\cal B},\quad a+a'-(b+b')>{\cal A}-{\cal B},& \hbox{region
II}\\
\hbox{otherwise},& \hbox{region III}
\end{cases}
\end{equation}
where
\begin{align}\label{43}
{\cal A}&:=u+s-\sqrt{|(u-\bar u)(v-\bar v)|}=\bar u+\bar
s-\sqrt{|(u-\bar u)(v-\bar v)|},\nonumber\\
{\cal B}&:=v+w-\sqrt{|(u-\bar u)(v-\bar v)|}=\bar v+\bar
w-\sqrt{|(u-\bar u)(v-\bar v)|}.
\end{align}
The relaxation time $\tau$ is then
\begin{equation}\label{44}
\tau =
\begin{cases}
[v+w+a+a'+(u-\bar u)(v-\bar v)(a+a'-u-s)^{-1}]^{-1},&\hbox{ region
I}\\ [u+s+b+b'+(u-\bar u)(v-\bar
v)(b+b'-v-w)^{-1}]^{-1},&\hbox{ region II}\\
\{v+w+u+s-2\textrm{Re}[\sqrt{(u-\bar u)(v-\bar v)}]\}^{-1},&\hbox{
region III}
\end{cases}
\end{equation}
It is seen that in region I, the relaxation time depends on the
injection and extraction rates at the beginning of the lattice; in
region II it depends on the injection and extraction rates at the
end of the lattice; and in region III it depends on none of the
injection or extraction rates.

The details of the calculations can be seen in \cite{AM2}.
\section{An asymmetric generalization of the Glauber model at zero
temperature, as an example} The Glauber model \cite{Gl} is a model
for the relaxation of an Ising model towards its equilibrium with
a heat bath at temperature $T$, which is based on the principle of
detailed balance. This model is based on a three-site interaction:
a spin is flipped with the rate $\mu:=1-\tanh\beta\, J$ if the
spins of its neighboring sites are the same as itself; it flips
with the rate $\lambda:=1+\tanh\beta\, J$, if the the spins of the
neighboring sites are opposite to it: and flips with the rate $1$,
if the spin of the neighboring sites are opposite to each other.
Here $\beta=(k_{\textrm{B}}\, T)^{-1}$, and time has been rescaled
so that one of the rates is $1$.

It has been shown in \cite{Sc2} that at zero temperature, the
Glauber dynamics is effectively a two-site interaction, in which
two opposite neighboring spins become the same, with the rate $1$
(independent of which is up and which is down, and which one
flips).

An extension of the Glauber dynamics at zero temperature is a
system with the following dynamics.
\begin{align}\label{45}
A\emptyset\to&\begin{cases}
              AA,&\hbox{ with the rate $u$}\\
              \emptyset\emptyset,&\hbox{ with the rate $v$}
              \end{cases}\nonumber\\
\emptyset A\to&\begin{cases}
              AA,&\hbox{ with the rate $v$}\\
              \emptyset\emptyset,&\hbox{ with the rate $u$}
              \end{cases}
\end{align}
where $A$ denotes a spin up (or a particle) and $\emptyset$
denotes a spin down (or a vacancy). In the ordinary glauber model,
the system is left-right symmetric and $u=v$. Now consider a
system with the dynamics \Ref{45}, on an $L$-site lattice. Assume
also that there are injection and extraction rates at the first
and last sites, with the rates $a$, $a'$, $b$, and $b'$ introduced
in the previous section. It is easily seen that for this system,
of the eight bulk-rate parameters introduced in the previous
section, only $u$, $v$, $\bar s$, and $\bar w$ are nonzero, and
\begin{align}\label{46}
\bar s&=u,\nonumber\\
\bar w&=v.
\end{align}
From this, it is seen that
\begin{align}\label{47}
x&=\frac{v}{u+v},\nonumber\\
y&=\frac{u}{u+v}.
\end{align}
So $x$ and $y$ are nonnegative and $x+y=1$. This means that there
is a static phase transition, which occurs at $u=v$, the point
corresponding to the ordinary Glauber model.

For the dynamical phase transition, it is seen that $\cal A$ and
$\cal B$ introduced in \Ref{43}, are
\begin{align}\label{48}
\cal A&:=u-\sqrt{u\, v},\nonumber\\
\cal B&:=v-\sqrt{u\, v}.
\end{align}
Of $\cal A$ and $\cal B$, only one can be positive, and this is
when $u\ne v$. So, as the injection and extraction rates are
nonnegative, at most two of the regions I, II, and III may exist:
for $u>v$, only the regions I and III; for $u<v$, only the regions
II and III. The expression for the relaxation time is
\begin{equation}\label{49}
\tau=
\begin{cases}
[v+a+a'+u\, v(a+a'-u)^{-1}]^{-1},&\hbox{ region I}\\
[u+b+b'+u\, v(b+b'-v)^{-1}]^{-1},&\hbox{ region II}\\
(v+u-2\sqrt{u\, v})^{-1},&\hbox{ region III}
\end{cases}
\end{equation}

As a final note, it is easily seen that changing the dynamics
\Ref{45} into the following
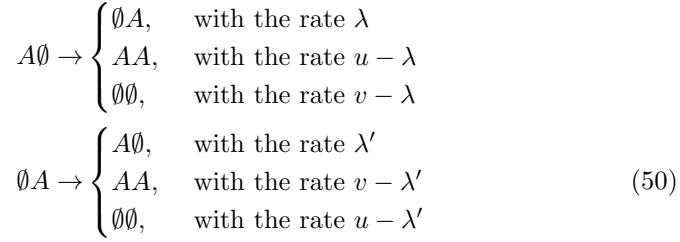
\begin{align}\label{50}
A\emptyset\to&\begin{cases}
              \emptyset A,&\hbox{ with the rate $\lambda$}\\
              AA,&\hbox{ with the rate $u-\lambda$}\\
              \emptyset\emptyset,&\hbox{ with the rate $v-\lambda$}
              \end{cases}\nonumber\\
\emptyset A\to&\begin{cases}
              A\emptyset,&\hbox{ with the rate $\lambda'$}\\
              AA,&\hbox{ with the rate $v-\lambda'$}\\
              \emptyset\emptyset,&\hbox{ with the rate $u-\lambda'$}
              \end{cases}
\end{align}
which contains diffusion as well, does not change the evolution
equation of the one-point function. So the results of this section
are valid for this interaction as well. The only difference is
that the introduction of diffusion, restricts the parameter space
of $u$ and $v$, since the rates should be nonnegative. So
$u,v\geq\lambda,\lambda'$.

\section{Extension to longer-range interactions: dynamical phase
transition of kinetic Ising model with boundaries} It was pointed
out in the previous section, that the Glauber dynamics at nonzero
temperatures consists of a three-site interaction. Let us rewrite
the reactions:
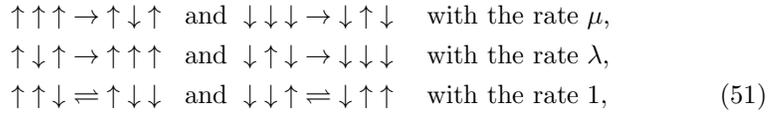
\begin{align}\label{51}
\uparrow \; \uparrow \; \uparrow \; \to \; \uparrow \; \downarrow
\;
 \uparrow \; \mbox{ and }
\downarrow \; \downarrow \; \downarrow \; \to \; \downarrow \;
\uparrow
\; \downarrow \; &\quad \mbox{with the rate } \mu, \nonumber \\
\uparrow \; \downarrow \; \uparrow \; \to \; \uparrow \; \uparrow
\; \uparrow \; \mbox{ and } \downarrow \; \uparrow \; \downarrow
\; \to
 \; \downarrow \;  \downarrow \;
\downarrow \; &\quad \mbox{with the rate } \lambda, \nonumber \\
\uparrow \; \uparrow \; \downarrow \; \rightleftharpoons \;
\uparrow \; \downarrow \; \downarrow \; \mbox{ and } \downarrow \;
\downarrow \; \uparrow \; \rightleftharpoons \; \downarrow \;
\uparrow \; \uparrow \; &\quad \mbox{with the rate } 1,
\end{align}
where an upward arrow means spin up, a downward arrow means spin
down, and $\lambda$ and $\mu$ are defined through
\begin{align}\label{52}
\lambda&:=1+\tanh\beta\, J,\nonumber\\
\mu&:=1-\tanh\beta\, J,
\end{align}
as in the previous section. As it is noted in \cite{Gl}, the
evolution equation for the expectation value of the spins (each
spin is either $1$ (upward) or $-1$ (downward)) is closed. This is
of course through for the spins at the bulk of the lattice. At the
boundaries, one should introduce some two-site interactions and
write the write separate equations for $\langle s_1\rangle$ and
$\langle s_L\rangle$. The new interactions introduced at the
boundaries are
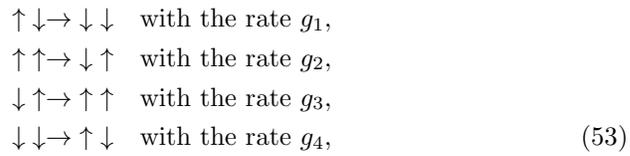
\begin{align}\label{53}
\uparrow \; \downarrow  \to \;  \downarrow \; \downarrow  &\quad
\mbox{with the rate } g_1, \nonumber \\
\uparrow \; \uparrow  \to \;  \downarrow \; \uparrow  &\quad
\mbox{with the rate } g_2, \nonumber \\
\downarrow \; \uparrow  \to \;  \uparrow \; \uparrow  &\quad
\mbox{with the rate } g_3, \nonumber \\
\downarrow \; \downarrow  \to \;  \uparrow \; \downarrow  &\quad
\mbox{with the rate } g_4,
\end{align}
For the spin flip of the first site, and
\begin{align}\label{54}
\downarrow \; \uparrow  \to \;  \downarrow \; \downarrow  &\quad
\mbox{with the rate } h_1, \nonumber \\
\uparrow \; \uparrow  \to \;  \uparrow \; \downarrow  &\quad
\mbox{with the rate } h_2, \nonumber \\
\uparrow \; \downarrow  \to \;  \uparrow \; \uparrow  &\quad
\mbox{with the rate } h_3, \nonumber \\
\downarrow \; \downarrow  \to \;  \downarrow \; \uparrow  &\quad
\mbox{with the rate } h_4,
\end{align}
For the spin flip of the last site. It can be seen that the
evolution equation for the one-point functions is closed, iff
\begin{align}\label{55}
g_1+g_4&=g_2+g_3,\nonumber\\
h_1+h_4&=h_2+h_3.
\end{align}
Provided this is through, the evolution equation is
\begin{align}\label{56}
\langle\dot s_i\rangle =&-2\langle s_i\rangle +
 (\langle s_{i+1}\rangle +\langle s_{i-1}\rangle )
 \tanh ({ 2\beta\, J}),\qquad 1<i<L \nonumber\\
\langle\dot s_1\rangle =& -(g_2+g_3)\langle s_1\rangle +
(g_1-g_2)\langle s_2\rangle
+(g_3-g_1),\nonumber\\
\langle\dot s_L\rangle =& -(h_2+h_3)\langle s_L\rangle +
(h_1-h_2)\langle s_{L-1}\rangle +(h_3-h_1).
\end{align}
The stationary solution for this is
\begin{equation}\label{57}
\langle s_j\rangle =D_1\,z_1^k+D_2\,z_2^{k-L-1},
\end{equation}
where
\begin{equation}\label{58}
z_1=z_2^{-1}=\tanh (\beta\, J).
\end{equation}
$D_1$ and $D_2$ are easily obtained at the thermodynamic limit
$L\to\infty$; and it is seen that they are both smooth functions
of the rates. So there is no static phase transition.

For the dynamical phase transition, similar to the method used in
the previous sections, one writes the eigenvalue problem for the
homogeneous equation. That equation reads
\begin{align}\label{59}
E\; X_j&= -2X_j+\tanh (2\beta J)(X_{j+1}+X_{j-1}),\quad
j\ne 1,L,\nonumber\\
E\; X_1&= -(g_2+g_3)X_1+(g_1-g_2)X_2,\nonumber\\
E\; X_L&= -(h_2+h_3)X_L+(h_1-h_2)X_{L-1}.
\end{align}
The solution to this is an expression like \Ref{35}, with
\begin{equation}\label{60}
E=-2+\tanh (2\beta\, J)(z_i+z_i^{-1}).
\end{equation}
and
\begin{align}\label{61}
z_i^{1-L}[2-g_2-g_3+z(g_1-g_2-\tanh (2\beta\, J))-z_i^{-1}\tanh
(2\beta\, J)] &\nonumber\\ \times[2-h_2-h_3+z_i(h_1-h_2-\tanh
(2\beta\, J))-z_i^{-1}\tanh (2\beta\, J)]\nonumber&\\ -
z_i^{L-1}[2-g_2-g_3+z^{-1}(g_1-g_2-\tanh
(2\beta\, J))-z_i\tanh (2\beta\, J)]&\nonumber\\
\times [2-h_2-h_3+z_i^{-1}(h_1-h_2-\tanh (2\beta\, J))-z_i\tanh
(2\beta\, J)]=0&.
\end{align}
Again in the thermodynamic limit $L\to\infty$, if all of the
solutions to the above equation are unimodular, the relaxation
time would be
\begin{equation}\label{62}
\tau=[-2+2\tanh(2\beta\, J)]^{-1},
\end{equation}
which is independent of the interaction rates at the boundaries.
If there are some solutions that are nonunimodolar, and among them
there is a solution with $\textrm{Re}(z_i+z_i^{-1})>2$ (for the
ferromagnetic case $J>0$), then the relaxation time becomes larger
than \Ref{62}. This is the slow phase. The criterion for this, is
that one of the roots of \Ref{61} passes the boundary
\begin{equation}\label{63}
y=\pm(x-1)\sqrt{\frac{x}{2-x}},
\end{equation}
where $x$ and $y$ are the real and imaginary parts of that root,
respectively.

In the thermodynamic limit $L\to\infty$, the equation \Ref{61} for
roots with modulus greater than one simplifies into
\begin{align}\label{64}
\{2-g_2-g_3+&z^{-1}[g_1-g_2-\tanh (2\beta\, J)]-z\tanh (2\beta\, J)\}\nonumber\\
\times \{2-h_2-h_3+&z^{-1}[h_1-h_2-\tanh (2\beta\, J)]-z\tanh
(2\beta\, J)\}=0.
\end{align}
So the criterion for the transition to occur, is that a complex
number $z=x+i y$, satisfies \Ref{63} and \Ref{64}. The relation
between rates, for this to occur, can be seen to be
\begin{align}\label{65}
&2[1-\tanh(2\beta\, J)]-g_2-g_4=0,\qquad \hbox{or},\nonumber\\
&[4\tanh(2\beta\, J)-2+g_2+g_3][g_1-g_2-\tanh(2\beta\,
J)]+(2-g_2-g_3)\tanh(2\beta\, J)=0.
\end{align}
This is for the case when the first factor in \Ref{64} vanishes. A
similar criterion come from the vanishing of the second factor,
with the roles of $g_i$'s and $h_i$'s interchanged.

The detailed calculations can be found in \cite{MA2}.
\section{Extension to multi-species systems: phase transition in an
asymmetric generalization of the zero-temperature Potts model} A
simple extension of the asymmetric generalization of the Glauber
model at zero temperature, in troduced in section 4, is an
asymmetric generalization of the zero-temperature Potts model. In
this system, any site can have $p+1$ states (rather than the two
states of the Glauber model). The bulk reactions are written as
\begin{equation}\label{66}
A_\alpha A_\beta\to\begin{cases}
                               A_\alpha A_\alpha,&\hbox{ with the
                               rate $u$}\nonumber\\
                               A_\beta A_\beta,&\hbox{ with the
                               rate $v$}
\end{cases}.
\end{equation}
Compare this with \Ref{45}. One adds to this, reaction rates at
the boundaries:
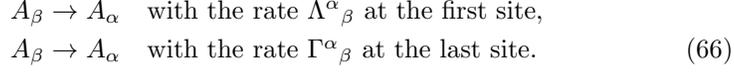
\begin{align}\label{67}
A_\beta\to A_\alpha &\quad\hbox{with the rate
$\Lambda^\alpha{}_\beta$ at the first site,}\nonumber\\
A_\beta\to A_\alpha &\quad\hbox{with the rate
$\Gamma^\alpha{}_\beta$ at the last site.}
\end{align}
For $\alpha\ne\beta$, $\Lambda^\alpha{}_\beta$ and
$\Gamma^\alpha{}_\beta$ are rates, and should be nonnegative. The
diagonal elements of $\Lambda$ and $\Gamma$ are chosen so that
\begin{equation}\label{68}
\mathbf{s}\,\Lambda=\mathbf{s}\,\Gamma=0.
\end{equation}
Using these, the evolution equation for the one-point functions is
written as
\begin{align}\label{69}
\langle \dot{\mathbf{n}}_j \rangle &= -(u+v)\langle
\mathbf{n}_j\rangle +u \langle\mathbf{n}_{j-1}\rangle + v\langle
\mathbf{n}_{j+1}\rangle,\quad j\ne 1,L\nonumber\\
\langle \dot{\mathbf{n}}_1\rangle &=\Lambda\langle
\mathbf{n}_1\rangle -v \langle\mathbf{n}_{1}\rangle + v\langle
\mathbf{n}_{2}\rangle,\nonumber \\
\langle \dot{\mathbf{n}}_L\rangle
=\Gamma\langle\mathbf{n}_L\rangle -u \langle\mathbf{n}_L\rangle +
u\langle\mathbf{n}_{L-1}\rangle.
\end{align}
\subsection{The static phase transition}
The stationary solution to \Ref{69} can be written as
\begin{equation}\label{70}
\langle\mathbf{n}_j\rangle=\mathbf{B}+\mathbf{C}'\,\left(\frac{u}{v}\right)^{j-1},
\end{equation}
or
\begin{equation}\label{71}
\langle\mathbf{n}_j\rangle=\mathbf{B}+\mathbf{C}''\,\left(\frac{u}{v}\right)^{j-L},
\end{equation}
where
\begin{align}\label{72}
\mathbf{s}\,\mathbf{B}&=1,\nonumber\\
\mathbf{s}\,\mathbf{C}'=\mathbf{s}\,\mathbf{C}''&=0,
\end{align}
and $\mathbf{B}$ and $\mathbf{C}'$ (or $\mathbf{C}''$) must also
satisfy conditions coming from the evolution equation \Ref{69} at
the boundaries. For $u<v$, it is better to work with \Ref{70}.
Then it is seen that in the thermodynamic limit $L\to\infty$, one
has
\begin{align}\label{73}
\Gamma\,\mathbf{B}&=0,\nonumber\\
(\Lambda-v+u)\mathbf{C}'&=-\Lambda\,\mathbf{B}.
\end{align}
The first equation has at least one non-zero solution for
$\mathbf{B}$, as $\gamma$ does have a left eigenvector
$\mathbf{s}$, with the eigenvalue zero. Depending on the
degeneracy of this zero eigenvalue, there are one or several
solutions for $\mathbf{B}$. One should then put $\mathbf{B}$ in
the second equation to find $\mathbf{C}'$. The conditions on
$\Lambda$, ensure that the real part of the eigenvalues of
$\Lambda$ are nonpositive, so $v-u$ cannot be an eigenvalue of
$\Lambda$. Hence, the second equation has one and only one
solution for $\mathbf{C}'$ (corresponding to each solution for
$\mathbf{B}$).

For $u>v$, one uses \Ref{71} and (in the thermodynamic limit)
arrives at
\begin{align}\label{74}
\Lambda\,\mathbf{B}&=0,\nonumber\\
(\Gamma-u+v)\mathbf{C}''&=-\Gamma\,\mathbf{B}.
\end{align}
Similar arguments hold for the solution of this equation, with the
roles of $\Lambda$ and $\Gamma$, $u$ and $v$, and $\mathbf{C}'$
and $\mathbf{C}''$ interchanged. Now apart from the question of
the uniqueness of the stationary profile (which depends on the
degeneracy of the matrices $\Gamma$ or $\Lambda$) one can see that
if $u<v$, the profile is flat near the last site, while for $u>v$
it is flat near the first sites. This is the static phase
transition, similar to what observed for the asymmetric
generalization of the zero-temperature Glauber model.
\subsection{The dynamical phase transition}
The eigenvalue equation for the homogeneous part of \Ref{69} is
\begin{align}\label{75}
E\,\mathbf{X}_j=&-(u+v)\mathbf{X}_j +u\,\mathbf{X}_{k-1}+v\,
\mathbf{X}_{k+1},\qquad j\ne 1,L\nonumber \\
E\,\mathbf{X}_1=& \Lambda\,\mathbf{X}_1 -v\,\mathbf{X}_1+v\,
\mathbf{X}_2,\nonumber \\
E\,\mathbf{X}_L=& \Gamma\,\mathbf{X}_L -u\,\mathbf{X}_L+u\,
\mathbf{X}_{L-1},
\end{align}
the solution to which is
\begin{equation}\label{76}
\mathbf{X}_j=\mathbf{b}\,z_1^j+\mathbf{c}\,z_2^j,
\end{equation}
where $z_i$'s satisfy
\begin{equation}\label{77}
E=-(u+v)+v\, z_i+u\,z_i^{-1}.
\end{equation}
The second and third equation of \Ref{75}, can be witten as
\begin{equation}\label{78}
  \begin{pmatrix}
    -(u+\Lambda )Z+\sqrt{u\,v} & -(u+\Lambda )Z^{-1}+\sqrt{u\,v} \\
                             &\\
    -(v+\Gamma )Z^L+\sqrt{u\,v}\,Z^{L+1} & -(v+\Gamma )Z^{-L}+\sqrt{u\,v}\,Z^{-L-1}
  \end{pmatrix}
  \begin{pmatrix}
    \mathbf{b} \\
    \\
    \mathbf{c}
  \end{pmatrix}
=0,
\end{equation}
where the variable
\begin{equation}\label{79}
Z_i:=z_i\sqrt{\frac{v}{u}}
\end{equation}
has been used, in terms of which
\begin{equation}\label{80}
E=-(u+v)+\sqrt{u\,v}(Z_i+Z_i^{-1}).
\end{equation}
\Ref{78} has nontrivial solutions for $\mathbf{b}$ and
$\mathbf{c}$, iff the determinant of the matrix of the
coefficients vanishes:
\begin{equation}\label{81}
\det\begin{pmatrix}
    -(u+\Lambda )Z+\sqrt{u\,v} & -(u+\Lambda )Z^{-1}+\sqrt{u\,v} \\
                             &\\
    -(v+\Gamma )Z^L+\sqrt{u\,v}\,Z^{L+1} & -(v+\Gamma )Z^{-L}+\sqrt{u\,v}\,Z^{-L-1}
  \end{pmatrix}
=0.
\end{equation}
Again, if in the thermodynamic limit $L\to\infty$, for all of the
solutions of the above equation $\mathrm{Re}(Z+Z^{-1})\leq 2$, the
relaxation time is independent of the reaction rates at
boundaries, and the system is in the fast phase. In order that the
system be in the slow phase, it is necessary (but not sufficient)
that at least one of the roots of \Ref{81} has modulus larger than
one. For this solution, \Ref{81} is simplified in the
thermodynamic limit to
\begin{equation}\label{82}
  \det [\sqrt{u\,v}\,Z-(v+\Gamma )]\det[\sqrt{u\,v}\,Z-(u+\Lambda )]=0.
\end{equation}
So, denoting the eigenvalues of $\Lambda$ and $\Gamma$ with
$\lambda$ and $\gamma$ respectively, This root must be
\begin{equation}\label{83}
Z=\frac{v+\gamma}{\sqrt{u\,v}}, \quad \hbox{or} \quad
Z=\frac{u+\lambda}{\sqrt{u\,v}}.
\end{equation}
The real parts of the eigenvalues $\lambda$ and $\gamma$ are
nonpositive. It is then seen that if $u\geq v$, then the real part
of $Z$ in the first case is not greater than $1$, and
$\mathrm{Re}(Z+Z^{-1})\leq 2$. So in this case, the only relevant
equation for finding the system in the slow phase is the
$Z=(u+\lambda)/\sqrt{u\, v}$. A similar argument shows that if
$v\geq u$, then the first equation of \Ref{83} is relevant. So,
without loss of generality, let's take $u<v$. In this case,
$\mathbf{b}$ should be an eigenvector of $\Gamma$. However, one
also demands
\begin{equation}\label{84}
\mathbf{s}\,\mathbf{b}=0.
\end{equation}
If $\mathbf{b}$ is the only eigenvector of $\Gamma$, with zero
eigenvalue, then \Ref{84} cannot be satisfied, as $\mathbf{s}$
would also the only left eigenvector of $\Gamma$ with zero
eigenvalue. So, from the eigenvalues of $\Gamma$, one should put
aside one zero eigenvalue, and consider only the other ones. Of
course if the zero eigenvalue is degenerate, then $\gamma$ can
still be zero.

The system undergoes a dynamical phase transition, when $x$ and
$y$ (the real and imaginary parts of $Z$) pass the curve
\begin{equation}\label{85}
y=\pm(x-1)\sqrt{\frac{x}{2-x}},\qquad x\geq 1.
\end{equation}
In terms of the eigenvalues of $\Gamma$, it is seen that the
system is in the slow phase iff
\begin{equation}\label{86}
|\mathrm{Im}(\lambda)|<\left[\mathrm{Re}(\lambda)+u-\sqrt{u\,v}\right]
\sqrt{\frac{\mathrm{Re}(\lambda)+u}{2\sqrt{u\,v}-\mathrm{Re}(\lambda)-u}},\hbox{
or}\quad\mathrm{Re}(\lambda)>2\sqrt{u\,v}-u.
\end{equation}

A simple way to induce the phase transition is to multiply the
matrix $\Gamma$ by a parameter $r$. This means multiplying the
rates of the reaction at the first site by $r$. As
$\mathrm{Re}(\gamma)\leq 0$, one can see that for a large enough
value of $r$, the value of $\mathrm{Re}(\gamma)+u-\sqrt{u\,v}$
will be negative (provided $\mathrm{Re}(\gamma)\ne 0$, that is,
provided the zero eigenvalue of the matrix $\Gamma$ is not
degenerate). So the system will be in the fast phase. It is also
seen that as $r$ tends to zero, either
$2\sqrt{u\,v}-\mathrm{Re}(\gamma)-u$ becomes negative, or in the
first inequality in \Ref{86} the right-hand becomes greater than
the left-hand side (which tends to zero). So, the system will be
in the slow phase. Roughly speaking, increasing the reaction rates
at the boundaries, brings the system from the slow phase
(relaxation time depending on the reaction rates at boundaries) to
the fast phase (relaxation time independent of the reaction rates
at boundaries). A similar argument holds, of course, for the case
$u>v$ and the eigenvalues of the matrix $\Lambda$.

The details of the calculations can be found in \cite{MAM}.

\end{document}